\documentclass[amsmath,amssymb,
aps,
nofootinbib,
prd,
superscriptaddress,
tightenlines,
notitlepage,
twocolumn,
showpacs,
floatfix]{revtex4-2}
\usepackage[utf8]{inputenc}
\usepackage{color}
\usepackage{xcolor}
\usepackage{hyperref}
\usepackage{mathtools}
\usepackage{graphicx}
\usepackage{dcolumn}
\usepackage{bm}

\newcommand{\be}{\begin{equation}}
\newcommand{\ee}{\end{equation}}
\newcommand{\bea}{\begin{equation}\begin{aligned}}
\newcommand{\eea}{\end{aligned}\end{equation}}

\newcommand{\IHEP}{\affiliation{State Key Laboratory of Particle Astrophysics, Institute of High Energy Physics,
Chinese Academy of Sciences, Beijing 100049, China}}

\newcommand{\hh}{\affiliation{School of Physics and Electromechanical Engineering, Hubei University of Education, Wuhan 430205, China}}

\begin{document}


\title{Axion Star Bosenova in Axion Miniclusters}
\author{Zihang Wang}
\email{wangzihang@hue.edu.cn}
\hh
\author{Yu Gao}
\email{gaoyu@ihep.ac.cn}
\IHEP
\date{\today}
\begin{abstract}
Axionic dark matter can form structures known as miniclusters that host an axion star at their center. The axion star feeds on the host, and the axion star mass may grow beyond its stability limit, leading to a potential bosenova. Since a dilute axion star has a stable mass limit only when self-interaction is considered, we include axion self-interaction effects in this paper, and specify the condition for bosenova in the QCD axion and temperature-independent axion-like particle parameter spaces. We find that self-interaction may dominate the mass growth of the axion star. For a minicluster with a large initial overdensity, bosenova occurs in a large fraction of axion parameter space. For the QCD axion, bosenova occurs within the age of the Universe for miniclusters with an initial overdensity $\delta\gtrsim 100$.
\end{abstract}

\maketitle

\section{Introduction}
It is known that cold dark matter comprises a significant part of the Universe. One of the most motivated low-mass candidates is the axion, which was originally proposed to solve the strong CP problem~\cite{Peccei:1977hh,Weinberg:1977ma,Wilczek:1977pj}. It also accounts for the relic abundance of the Universe~\cite{Sikivie:1982qv,Preskill:1982cy,Abbott:1982af,Dine:1982ah,Berezhiani:1992rk} through the misalignment mechanism. In addition, generalized axion-like particles (ALPs) are also well motivated, e.g. in string theory~\cite{Svrcek:2006yi} by dimension compatification, as well as in many other new physics models~\cite{DiLuzio:2020wdo}. 

The low axion mass naturally arises as a pseudo-Nambu-Goldstone boson of a spontaneously broken Peccei-Quinn (PQ) symmetry. If this symmetry breaking occurs after the end of inflation, the axion field is expected to vary from one horizon to another. Primordial fluctuations will grow as the Universe expands, and form dense regions known as miniclusters (MCs)~\cite{Hogan:1988mp,Kolb:1994fi,Chang:1998tb}. These MCs may further aggregate to form minicluster halos (MCHs)~\cite{Fairbairn:2017sil,Xiao:2021nkb}. Later, axion stars can form inside MCs and MCHs. 

The mass of axion MCs is determined by the axion mass contained in the Hubble horizon when the axion field starts to oscillate~\cite{Enander:2017ogx}. The size of axion MCs depends on the initial overdensity $\delta$. Compact axion MCs and MCHs may survive tidal disruptions and remain until present~\cite{Fairbairn:2017sil,Kavanagh:2020gcy,DSouza:2024flu,DSouza:2024uud}. Such MCs can be studied or constrained by microlensing~\cite{Fairbairn:2017sil}, pulsar timing arrays~\cite{Siegel:2007fz,Lee:2020wfn}, wide binary evaporation~\cite{Qiu:2024muo,Wang:2024pqa}, etc. Equilibrium between quantum pressure, gravity, and axion self-interaction leads to stable solutions, a.k.a. axion stars~\cite{Kolb:1993zz,Chavanis:2011zi,Chavanis:2011zm,Schiappacasse:2017ham,Visinelli:2017ooc}. They may naturally form and grow within an isolated MC~\cite{Eggemeier:2019jsu}, MCHs~\cite{Chang:2024fol}, minihalos around primordial black holes~\cite{Hertzberg:2020hsz,Yin:2024xov} or form directly at matter-radiation equality~\cite{Gorghetto:2024vnp}. Previous works have shown that axion stars form in an axion MC after a characteristic time $\tau$, after which the axion star continues to grow. Such a picture is confirmed in numerical simulations~\cite{Levkov:2018kau,Dmitriev:2023ipv}. The mass growth rate can also be obtained from analytical considerations, by considering the capture rate of axions by the axion star~\cite{Chan:2022bkz}. The similar calculations can also be used to obtain the capture rate of axions by stars~\cite{Budker:2023sex}. This raises the question of whether an axion star grows beyond the maximal stable mass. If this maximum is reached, the equilibrium state of the dilute axion star no longer exists, and it will quickly collapse and explode as a bosenova via run-away self-interaction~\cite{Levkov:2016rkk}. Such a violent process may be associated with observable signals, and even studied as a potential standard candle~\cite{Di:2024tlz}. It has also been argued that such a process may lead to a decrease in the amount of cold dark matter, and the corresponding axion parameters are constrained~\cite{Fox:2023xgx}.

The axion star will have a maximum mass and become a bosenova only with self-interaction.
In this work, we focus on the axionic self-interaction's role in the process of a dilute axion star reaching its maximum mass, and subsequent de-stabilization. This provides a more concrete study of the growth in addition to gravitational effects studied in Ref.~\cite{Chan:2022bkz}. The mass growth rate may be dominated by self-interaction for certain axion parameters. We will determine the condition and the timescale for the process in terms of axion properties. While the initial axion star formation in an MC still needs numerical simulation, the star growth can be treated analytically. Our results show that for a wide range of axion parameters, including the QCD axion case, the bosenova will occur. QCD axion bosenova occurs in the present Universe for MCs with an initial overdensity $\delta\gtrsim 100$. In the ALP case with a temperature-independent mass, bosenova occurs for $\delta\gtrsim 1$, and can be faster than the QCD axion case. 
Bosenovae can potentially associate with electromagnetic and gravitational wave signals~\cite{Escudero:2023vgv,Di:2023nnb}, and leads to a change in the amount of radiation and dark matter in the Universe~\cite{Fox:2023xgx}.

This paper is organized as follows: In Section~\ref{sec:a}, we present the calculation for the mass growth rate of an axion star in an axion MC, considering both gravity and self-interaction. In Section~\ref{sec:b}, we consider realistic parameters of axion MCs and MCHs to obtain the time that the axion star reaches the maximum mass and becomes a bosenova. We further consider ALPs in Section~\ref{sect:alp}, then summarize and conclude in Section~\ref{sect:diss}.

\section{Axion star growth in a minicluster}
\label{sec:a}
In this section, we will calculate the axion star mass growth rate in an MC.
We will briefly review the formation and the mass and density of MCs first. In the early Universe, the axion field $\varphi$ evolves as
\begin{equation}
\ddot{\varphi}+3H(t)\dot{\varphi}+m_{a}^{2}(t)\varphi=0\, ,
\end{equation}
where $m_{a}(t)$ is the axion mass that may depends on temperature~\cite{GrillidiCortona:2015jxo}. At time $t_{\rm osc}$, the axion field overcomes Hubble friction and starts to oscillate, determined by $m_{a}(t_{\rm osc})=3H(t_{\rm osc})$. The characteristic mass $M_{0}$ of a typical MC is given by the mass contained inside the Hubble horizon at $t_{\rm osc}$:
\begin{equation}\label{eq:m0}
M_{0}=\frac{4\pi^4}{3}\frac{\bar{\rho}_{a}(t_{\rm osc})}{H^{3}(t_{\rm osc})}\, ,
\end{equation}
where $\bar{\rho}_{a}(t_{\rm osc})$ and $H(t_{\rm osc})$ are the average density of axions and the Hubble parameter at $t_{\rm osc}$. The MC mass $M_{\rm MC}$ is of order $M_{0}$, and their densities depend on the initial overdensity $\delta$. The MC density is~\cite{Kolb:1994fi}
\begin{equation}\label{eq:MCd}
\rho_{\rm MC}=140\,\delta^{3}(1+\delta)\bar{\rho}_{a}(z_{\rm eq})\, ,
\end{equation}
where $\bar{\rho}_{a}(z_{\rm eq})$ is the average density of axions at matter-radiation equality. It can be seen that the MC density depends sensitively on the overdensity $\delta$. The MC virializes, and its typical velocity is

\begin{equation}\label{eq:MCv}
v_{g}\sim\sqrt{\frac{GM_{\rm MC}}{R_{\rm MC}}}\, ,
\end{equation}
where $R_{\rm MC}$ is a characteristic radius of MC. The momentum distribution of axions in an MC is taken as Maxwellian~\cite{Chan:2022bkz},
\begin{equation}\label{eq:momentum}
f_{\vec{k}}=f_{g}e^{-\frac{\vec{k}^{2}}{k_{g}^{2}}}
\end{equation}
where $f_{\vec{k}}$ is the occupation number for momentum $\vec{k}$, and we take $k_{g}=m_{a}v_{g}$. The momentum distribution is normalized so that
\begin{equation}
\int\frac{\textrm{d}^{3}\vec{k}}{(2\pi)^{3}}f_{\vec{k}}=n_{a}=\frac{\rho_{\rm MC}}{m_{a}}\, .
\end{equation}
The normalization factor $f_{g}$ is
\begin{equation}
f_{g}=(4\pi)^{\frac{3}{2}}\frac{\rho_{\rm MC}}{m_{a}k_{g}^{3}}\, .
\end{equation}
We can verify that the occupation number $f_{g}\gg 1$.
The momentum distribution $f_{\vec{k}}$ evolves as the axion star grows. The change of characteristic momentum $k_{g}$ is negligible as long as the axion star mass is much smaller than the MC mass. Although the distribution may evolve, the mass growth of the axion star is insensitive to the specific form of momentum distribution as long as $k_{g}$ is unchanged. Hence, in the following we will assume Eq.~(\ref{eq:momentum}) as the axion star grows.

We will consider an axion star in the center of an MC. Numerical simulations show that such an axion star can naturally form within an MC~\cite{Levkov:2018kau,Eggemeier:2019jsu,Dmitriev:2023ipv}. The axion star appears after a timescale $\tau_{c}$. Before $\tau_{c}$, the momentum distribution within the MC changes gradually, with an accumulation of low momentum modes~\cite{Levkov:2018kau}. After the axion star is formed, it will continue to absorb axions and grow. The growth due to gravitational interaction has been calculated in Ref.~\cite{Chan:2022bkz}.
However, the dilute axion star has a maximum mass only when self-interaction is involved, and self-interaction may dominate the mass growth. Hence we also need to consider axion star mass growth due to the axion self-interaction.

Next, we will calculate the mass growth rate of an axion star in an MC.
We will use a similar approach as in Ref.~\cite{Chan:2022bkz} to obtain the mass growth rate. We will regard the absorption of axions as a quantum transition process and calculate the absorption rate. 
The axion action is
\begin{equation}
S=\int \textrm{d}^{4}x\, \left(\frac{1}{2}\partial_{\mu}a\partial^{\mu}a-\frac{1}{2}m_{a}^{2}a^{2}-\frac{1}{24}\lambda a^{4}\right)\, .
\end{equation}
Here, we only write down the self-interaction part and omit the gravitational interaction terms. We work in the nonrelativistic limit, so that the axion field can be written as 
\begin{equation}
a(\vec{x},t)=\frac{1}{\sqrt{2m_{a}}}(e^{-im_{a}t}\psi(\vec{x},t)+e^{im_{a}t}\psi^{*}(\vec{x},t)) \, .
\end{equation}
where $e^{-im_a t}$ is factored out from the time part of the wavefunction, and nonrelativistically we expect $\partial_t \psi\ll im_a\psi$, so that the $\partial_{t}\psi\partial_{t}\psi^{*}$ term can be suppressed.
The nonrelativistic action then becomes
\begin{equation}
\begin{split}
S=\int \textrm{d}t\textrm{d}^{3}\vec{x}\, \left(i\psi^{*}\partial_{t}\psi+\frac{1}{2m_{a}}\psi^{*}\nabla^{2}\psi-\frac{\lambda}{16m_{a}^{2}}|\psi|^{4}\right)\, .
\end{split}
\end{equation}
Now we consider an axion star resides in an MC. The MC state is denoted as `g', while the state in the axion star is denoted as `s'. The axion field can be written as $\psi=\psi_{s}+\psi_{g}$. We will treat $\psi_{s}$ as a background field here. The action now becomes
\begin{equation}\label{eq:gs}
\begin{split}
&S_{g}=\\ &\int \textrm{d}t\textrm{d}^{3}\vec{x}\, \left(i\psi_{g}^{*}\partial_{t}\psi_{g}+\frac{1}{2m_{a}}\psi_{g}^{*}\nabla^{2}\psi_{g}-\frac{\lambda}{16m_{a}^{2}}|\psi_{g}+\psi_{s}|^{4}\right)\, .
\end{split}
\end{equation}
where the field equation of $\psi_{s}$ (including gravity) can be used to eliminate derivative terms involving $\psi_{s}$.
Now we write down the axion star wave function, and separate out a time-dependent factor,
\begin{equation}
\psi_{s}(\vec{x},t)=\chi(r)e^{-i\epsilon_{s}t} \, .
\end{equation}
We consider the case that the axion star is spherically symmetric, hence the spatial part only depends on $r\equiv |\vec{x}|$. The form of $\chi(r)$ can be taken as a real function,
\begin{equation}\label{eq:profile}
\chi(r)=A\left(1+\frac{r^{2}}{R^{2}}\right)^{-4} \, ,
\end{equation}
where $R$ is the characteristic radius of the axion star, $A$ is a normalization constant, 
\begin{equation}
A=\sqrt{\frac{N}{c_{1}R^{3}}} \, .
\end{equation}
where $N$ is the number of axions in the axion star, and $c_{1}=0.318$.

The wave function of the MC can be written as a sum of waves from all directions,
\begin{equation}\label{eq:psig}
\psi_{g}(\vec{x},t)=\frac{1}{\sqrt{V}}\sum_{\vec{k}} a_{\vec{k}}\varphi_{\vec{k}}(\vec{x})e^{-i\epsilon_{\vec{k}} t} \, ,
\end{equation}
where $\epsilon_{\vec{k}}=k^{2}/(2m_{a})$ is the kinetic energy. $\{\varphi_{\vec{k}}(\vec{x})\}$ is a suitable basis and $a_{\vec{k}}$ is its expansion coefficient. After second quantization, we will promote $a_{\vec{k}}$ to an annihilation operator. If gravity is negligible, $\varphi_{\vec{k}}(\vec{x})$ can be taken as plane waves. However, the gravity of the axion star must be taken into account here, because the MC axion velocity around the axion star can be much larger than the characteristic MC axion velocity $v_{g}$.

The gravitational potential has the same form as the Coulomb potential. We use scattering-state solutions~\cite{qm} of the Coulomb potential as the basis $\varphi_{\vec{k}}(\vec{x})$. The momentum $\vec{k}$ here represents axion momentum far away from the axion star. Define dimensionless parameters $\vec{\kappa}=\vec{k}R$ and $\vec{\xi}=\vec{x}/R$.
The form of $\varphi_{\vec{\kappa}}$ is
\begin{equation}\label{eq:1F1}
\varphi_{\vec{\kappa}}\left(\vec{\xi}\right)=e^{i\vec{\kappa}\cdot\vec{\xi}}\,\Gamma\left(1-\frac{i\beta}{\kappa}\right)\, e^{\frac{\pi\beta}{2\kappa}}  \prescript{}{1}{F}_{1}\left[\frac{i\beta}{\kappa},1,i\left(\kappa \xi-\vec{\kappa}\cdot\vec{\xi}\right)\right] \, .
\end{equation}
where $\beta=Gm_{a}^{2}M_{s}R$ is a constant related to the Coulomb potential, $\xi\equiv |\vec{\xi}|$ and $M_{s}$ is the mass of the axion star. The above wave function has a simpler form for $\kappa\ll 2\pi\beta$ and $\xi\lesssim 2\pi/\kappa$~\cite{Budker:2023sex},
\begin{equation}\label{eq:J0}
\varphi_{\vec{\kappa}}\left(\vec{\xi}\right)=\sqrt{\frac{2\pi\beta}{\kappa}}J_{0}\left[2\sqrt{\beta\left(\xi-\vec{n}\cdot\vec{\xi}\right)}\right] \, .
\end{equation}
The plane wave far away from the axion star is distorted to the form Eq.~(\ref{eq:J0}) close to the axion star with $\xi\lesssim 2\pi/\kappa$.
Next we consider absorption of axions by the axion star. The axions in the MC (state `g') must be absorbed to become the axion star's `s' state. 
The kinetic energy of axions in an MC $\epsilon_{g}$ is positive, while the binding energy $\epsilon_{s}<0$. The only two-body processes that conserve energy and contribute to axion star mass growth is $g+g\rightarrow g+s$ and its inverse process $g+s\rightarrow g+g$. The former is an absorption process where two axions in the gas state interact and one of them is captured by the axion star. The latter is the reverse process which will evaporate the axion star. For axion stars that are relatively heavy, as the case in this paper, absorption always dominates over evaporation.

We expand the last term in the Lagrangian Eq.~(\ref{eq:gs}) and only keep terms that correspond to the process $g+g\rightarrow g+s$ and $g+s\rightarrow g+g$. The only interaction term that contributes to the process is,
\begin{equation}
\mathcal{L}_{\rm int}=-\frac{\lambda}{8m_{a}^{2}}|\psi_{g}|^{2}\psi_{s}^{*}\psi_{g}+h.c. \, .
\end{equation}
The interaction Hamiltonian is
\begin{equation}
\mathcal{H}_{\rm int}=\frac{\lambda}{8m_{a}^{2}}|\psi_{g}|^{2}\psi_{s}^{*}\psi_{g}+h.c. \, .
\end{equation}
We treat $\psi_{s}$ as a classical solution while $\psi_{g}$ as quantized. Now $a_{\vec{k}}$ in Eq.~(\ref{eq:psig}) becomes an annihilation operator. First-order perturbation theory gives a transition amplitude
\begin{equation}
\mathcal{M}=-i\int \textrm{d}^3{x}\textrm{d}t \,\langle{f}| \mathcal{H}_{\rm int} |i \rangle \, .
\end{equation}
The number of reaction $g_{1}+g_{2}\rightarrow g_{3}+s$ is
\begin{equation}\label{eq:N0}
N_{0}=\frac{V^3}{2}\int\frac{\textrm{d}^{3}\vec{k}_{1}}{(2\pi)^{3}}\frac{\textrm{d}^{3}\vec{k}_{2}}{(2\pi)^{3}}\frac{\textrm{d}^{3}\vec{k}_{3}} {(2\pi)^{3}} \left|\mathcal{M}_{\vec{k}_{1}+\vec{k}_{2}\rightarrow\vec{k}_{3}+s}\right|^{2} \, ,
\end{equation}
where a factor $1/2$ is introduced to avoid double counting. The volume $V$ is the same normalization factor as in Eq.~(\ref{eq:psig}). We will consider the absorption process first. The momentum of $g_{1},g_{2},g_{3}$ is $\vec{k}_{1},\vec{k}_{2},\vec{k}_{3}$, and the corresponding occupation number is $f_{1},f_{2},f_{3}$. The square of the amplitude of the absorption process is 
\begin{equation} 
\begin{split}
\left|\mathcal{M}_{g+g\rightarrow g+s}\right|^{2}&=\left|\int \textrm{d}^3{x}\textrm{d}t \,\langle{f}| \mathcal{H}_{\rm int} |i \rangle\right|^{2}\\ &=\frac{\lambda^{2}}{64m_{a}^{4}}\left|\int \textrm{d}^3{x}\textrm{d}t \,\langle{f}| \psi_{s}^{*}|\psi_{g}|^{2}\psi_{g} |i \rangle\right|^{2}
\end{split}
\end{equation}
The matrix element inside the integral is
\begin{equation} 
\begin{split}
&\langle{f}|\psi_{s}^{*} |\psi_{g}|^{2}\psi_{g}|i \rangle=\\ &V^{-\frac{3}{2}}\sum_{\vec{k}'_{1}\vec{k}'_{2}\vec{k}'_{3}} \langle{f_{1}-1, f_{2}-1, f_{3}+1}|a_{\vec{k}'_{1}}a_{\vec{k}'_{2}}a^{\dagger}_{\vec{k}'_{3}}|f_{1},f_{2},f_{3} \rangle \cdot \\ & \varphi_{\vec{k}'_{1}}(x)\varphi_{\vec{k}'_{2}}(x)\varphi^{*}_{\vec{k}'_{3}}(x) \chi(r) e^{i(\epsilon'_{3}+\epsilon'_{s}-\epsilon'_{1}-\epsilon'_{2})t}
\end{split}
\end{equation}
Hence, the space-time integral becomes
\begin{equation} 
\begin{split}
&\int \textrm{d}^3{x}\textrm{d}t\, \langle{f}|\psi_{s}^{*} |\psi_{g}|^{2}\psi_{g}|i \rangle =\\ &V^{-\frac{3}{2}}\sum_{\vec{k}'_{1}\vec{k}'_{2}\vec{k}'_{3}}  \langle{f_{1}-1, f_{2}-1, f_{3}+1}|a_{\vec{k}'_{1}}a_{\vec{k}'_{2}}a^{\dagger}_{\vec{k}'_{3}}|f_{1},f_{2},f_{3} \rangle \cdot \\ &\left[\int \textrm{d}^3{x}\,\varphi_{\vec{k}'_{1}}(x)\varphi_{\vec{k}'_{2}}(x)\varphi^{*}_{\vec{k}'_{3}}(x) \chi(r)\right] \int \textrm{d}t\, e^{i(\epsilon'_{3}+\epsilon'_{s}-\epsilon'_{1}-\epsilon'_{2})t}
\end{split}
\end{equation}
where we write down the occupation number for momentum state $\vec{k}_{1},\vec{k}_{2},\vec{k}_{3}$ explicitly.
We use the relations $a^{\dagger}_{\vec{k}'_{3}}|f_{3}\rangle=\sqrt{f_{3}+1} \,\delta_{\vec{k}_{3},\vec{k}'_{3}}|f_{3}+1\rangle$,
$a_{\vec{k}'_{1}}|f_{1}\rangle=\sqrt{f_{1}}\,\delta_{\vec{k}_{1},\vec{k}'_{1}}|f_{1}-1\rangle$, 
$a_{\vec{k}'_{2}}|f_{2}\rangle=\sqrt{f_{2}} \,\delta_{\vec{k}_{2},\vec{k}'_{2}}|f_{2}-1\rangle$ to obtain
\begin{equation} 
\begin{split}
\int &\textrm{d}^3{x}\textrm{d}t\, \langle{f}|\psi_{s}^{*}|\psi_{g}|^{2}\psi_{g}|i \rangle =\\ &2\sqrt{f_{1}f_{2}(1+f_{3})}\,V^{-\frac{3}{2}} \left[\int \textrm{d}^3{x}\,\varphi_{\vec{k}_{1}}(x)\varphi_{\vec{k}_{2}}(x)\varphi^{*}_{\vec{k}_{3}}(x) \chi(r)\right] \\& \int \textrm{d}t\, e^{i(\epsilon_{3}+\epsilon_{s}-\epsilon_{1}-\epsilon_{2})t}
\end{split}
\end{equation}
A factor $2$ appears because there are two ways of pairings ($\vec{k}'_{1}\rightarrow \vec{k}_{1}$, $\vec{k}'_{2}\rightarrow \vec{k}_{2}$ and $\vec{k}'_{1}\rightarrow \vec{k}_{2}$, $\vec{k}'_{2}\rightarrow \vec{k}_{1}$). The square of the time integration can be written as 
\begin{equation}
 \left|\int \textrm{d}t\, e^{i(\epsilon_{3}+\epsilon_{s}-\epsilon_{1}-\epsilon_{2})t}\right|^{2}=(2\pi T)\delta(\epsilon_{1}+\epsilon_{2}-\epsilon_{3}-\epsilon_{s}) \, .
\end{equation}
where $T$ is the time duration considered. The number of reactions divided by $T$ is the reaction rate. The time integration gives an energy conservation delta function,
\begin{equation} 
\begin{split}
&\left|\int \textrm{d}^3{x}\textrm{d}t\, \langle{f}|\psi_{s}^{*}|\psi_{g}|^{2}\psi_{g}|i \rangle\right|^{2} = 4f_{1}f_{2}(1+f_{3}) V^{-3}\cdot \\ &\left|\int \textrm{d}^3{x}\,\varphi_{\vec{k}_{1}}(x)\varphi_{\vec{k}_{2}}(x)\varphi^{*}_{\vec{k}_{3}}(x) \chi(r)\right|^{2}  (2\pi T)\delta(\epsilon_{1}+\epsilon_{2}-\epsilon_{3}-\epsilon_{s}) 
\end{split}
\end{equation}
Now the number of reactions can be written as
\begin{equation} 
\begin{split}
&N_{0} = \frac{\lambda^{2}}{64m_{a}^{4}}\int\frac{\textrm{d}^{3}\vec{k}_{1}}{(2\pi)^{3}}\frac{\textrm{d}^{3}\vec{k}_{2}}{(2\pi)^{3}}\frac{\textrm{d}^{3}\vec{k}_{3}} {(2\pi)^{3}}\, 4f_{1}f_{2}(1+f_{3})\cdot T \\ &\left|\int \textrm{d}^3{x}\,\varphi_{\vec{k}_{1}}(x)\varphi_{\vec{k}_{2}}(x)\varphi^{*}_{\vec{k}_{3}}(x) \chi(r)\right|^{2}  (2\pi)\delta(\epsilon_{1}+\epsilon_{2}-\epsilon_{3}-\epsilon_{s}) 
\end{split}
\end{equation}
Dividing the number of reactions by $T$, we obtain the growth rate of the number of axions in the axion star,
\begin{equation}
\begin{split}
&\frac{\textrm{d}N}{\textrm{d}t}=\frac{\textrm{d}N_{0}}{\textrm{d}t} =\\ &\frac{\lambda^{2}}{64m_{a}^{4}}\int\frac{\textrm{d}^{3}\vec{k}_{1}}{(2\pi)^{3}}\frac{\textrm{d}^{3}\vec{k}_{2}}{(2\pi)^{3}}\frac{\textrm{d}^{3}\vec{k}_{3}} {(2\pi)^{3}}(2\pi)\delta(\epsilon_{1}+\epsilon_{2}-\epsilon_{3}-\epsilon_{s})
\\
& 4f_{1}f_{2}(1+f_{3}) \left|\int \textrm{d}^{3}\vec{x}\, \chi(r)\varphi_{\vec{k}_{1}}(x)\varphi_{\vec{k}_{2}}(x)\varphi^{*}_{\vec{k}_{3}}(x) \right|^{2} \, .
\end{split}
\end{equation}

The inverse process $s+g\rightarrow g+g$ can be calculated similarly. The main difference is that the matrix element becomes,
\begin{equation} 
\begin{split}
&\langle{f}||\psi_{g}|^{2}\psi_{s}^{*}\psi_{g}|i \rangle= \\ &V^{-\frac{3}{2}}\sum_{\vec{k}'_{1}\vec{k}'_{2}\vec{k}'_{3}} \langle f_{1}+1,f_{2}+1,f_{3}-1|a^{\dagger}_{\vec{k}'_{1}}a^{\dagger}_{\vec{k}'_{2}}a_{\vec{k}'_{3}}|f_{1},f_{2},f_{3}\rangle \cdot \\ & \varphi_{\vec{k}'_{1}}(x)\varphi_{\vec{k}'_{2}}(x)\varphi^{*}_{\vec{k}'_{3}}(x) \chi(r) e^{i(\epsilon'_{3}+\epsilon'_{s}-\epsilon'_{1}-\epsilon'_{2})t}
\end{split}
\end{equation}

Taking account of the inverse process $s+g\rightarrow g+g$, the growth rate of the number of axions is
\begin{equation}
\begin{split}
\frac{\textrm{d}N}{\textrm{d}t}&=\frac{\lambda^{2}}{64m_{a}^{4}}\int\frac{\textrm{d}^{3}\vec{k}_{1}}{(2\pi)^{3}}\frac{\textrm{d}^{3}\vec{k}_{2}}{(2\pi)^{3}}\frac{\textrm{d}^{3}\vec{k}_{3}} {(2\pi)^{3}}(2\pi)\delta(\epsilon_{1}+\epsilon_{2}-\epsilon_{3}-\epsilon_{s})\cdot
\\
& 4[f_{1}f_{2}(1+f_{3})-(1+f_{1})(1+f_{2})f_{3}] \cdot \\  & \left|\int \textrm{d}^{3}\vec{x}\, \chi(r)\varphi_{\vec{k}_{1}}(x)\varphi_{\vec{k}_{2}}(x)\varphi^{*}_{\vec{k}_{3}}(x) \right|^{2} \, .
\end{split}
\end{equation}
Because $f_{1},f_{2},f_{3}\gg 1$, the terms in the bracket $f_{1}f_{2}(1+f_{3})-(1+f_{1})(1+f_{2})f_{3}$ roughly equals to $f_{1}f_{2}-f_{1}f_{3}-f_{2}f_{3}$. \footnote{If we further treat the state `s' quantum mechanically and include the creation and annihilation operators, the factor becomes $f_{1}f_{2}-f_{1}f_{3}-f_{2}f_{3}+f_{1}f_{2}f_{3}/N$. The additional factor $f_{1}f_{2}f_{3}/N$ is small as long as $N\gg f_{1},f_{2},f_{3}$, which is satisfied for the QCD axion parameters discussed here and ALP MCs with $\delta\sim 1$. }  
The mass growth rate of a boson star is
\begin{equation}
\Gamma\equiv \frac{1}{M_{s}}\frac{\textrm{d}M_{s}}{\textrm{d}t}= \frac{m_{a}}{M_{s}} \frac{\textrm{d}N}{\textrm{d}t} \, ,
\end{equation}
where $M_{s}$ is the boson star mass. To evaluate the mass growth rate, we need to obtain the mass-radius relation of the axion star first. Then we can evaluate the binding energy $\epsilon_{s}$ and $\beta$ (related to the gravitational potential). 

The radius of dilute axion stars can be evaluated by a variational method~\cite{Schiappacasse:2017ham}. We calculate the total energy of the boson star using the profile Eq.~(\ref{eq:profile}), and find the radius $R$ that minimizes the total energy. The total energy includes kinetic energy, gravitational potential energy and self-interaction energy,
\begin{equation}
H=H_{\rm kin}+H_{\lambda}+H_{\rm grav} \, .
\end{equation}
We will use dimensionless parameters
\begin{equation}\label{eq:dimless}
\tilde{N}=Nm_{a}\sqrt{G|\lambda|}\, ,
\\ \tilde{R}=Rm_{a}^{2}\sqrt{\frac{G}{|\lambda|}}\, ,
\\ \tilde{H}=H\frac{\sqrt{|\lambda|^{3}}}{m_{a}^{2}\sqrt{G}}\, .
\end{equation}
This yields a dimensionless total energy,
\begin{equation}
\tilde{H}=4.337\frac{\tilde{N}}{\tilde{R}^{2}}-0.968\frac{\tilde{N}^{2}}{\tilde{R}}-0.0607\frac{\tilde{N}^{2}}{\tilde{R}^{3}} \, .
\end{equation}
For $\tilde{H}=a\frac{\tilde{N}}{\tilde{R}^{2}}-b\frac{\tilde{N}^{2}}{\tilde{R}}-c\frac{\tilde{N}^{2}}{\tilde{R}^{3}}$, the radius that minimize total energy is,
\begin{equation}\label{eq:radius}
\tilde{R}=\frac{a+\sqrt{a^{2}-3bc\tilde{N}^{2}}}{b\tilde{N}} \, .
\end{equation}
Here we note that $a^{2}-3bc\tilde{N}^{2}>0$, which provide an upper limit for $\tilde{N}$. The maximum $\tilde{N}$ allowed is $\tilde{N}_{\rm max}=10.33$. For a larger particle number $\tilde{N}$, the stable dilute axion star does not exist. This means that a dilute axion star has a maximum mass. If an axion star exceeds the mass limit, it will collapse and emit relativistic axions and possibly electromagnetic waves~\cite{Levkov:2016rkk}. The axion star that exceeds the mass limit will become a bosenova. 

Next, we determine the binding energy $\epsilon_{s}$. The axion star wave function is
\begin{equation}\label{eq:axst}
\psi_{s}(\vec{x},t)=\chi(r)e^{-i\epsilon_{s}t} \, .
\end{equation}
The solution $\psi_{s}$ satisfies the field equation,
\begin{equation}
i\partial_{t}\psi_{s}=-\frac{1}{2m_{a}}\nabla^{2}\psi_{s}+m_{a}\Phi\psi_{s}+\frac{\lambda}{8m_{a}^{2}}|\psi_{s}|^{2}\psi_{s} \, .
\end{equation}
Using Eq.(~\ref{eq:axst}) we obtain an equation for $\chi(r)$,
\begin{equation}\label{eq:chi}
\epsilon_{s}\chi=-\frac{1}{2m_{a}}\nabla^{2}\chi+m_{a}\Phi\chi+\frac{\lambda}{8m_{a}^{2}}|\chi|^{2}\chi \, .
\end{equation}
Multiply $\chi^{*}(r)$ 
from the left and integrate over $\vec{x}$, the terms on the right-hand side of Eq.~(\ref{eq:chi}) have the same form as $H_{\rm kin}$, $H_{\rm grav}$ and $H_{\lambda}$. We use the normalization condition
\begin{equation}
\int \textrm{d}^{3}x\, \psi^{*}_{s}(\vec{x},t) \psi_{s}(\vec{x},t)=\int \textrm{d}^{3}x\, \chi^{*}(r)\chi(r) =N \, .
\end{equation}
Then we obtain the binding energy,
\begin{equation}
\epsilon_{s}=\frac{1}{N}(H_{\rm kin}+2H_{\rm grav}+2H_{\lambda}) \, .
\end{equation}
For the approximate solution Eq.~(\ref{eq:profile}), the binding energy is
\begin{equation}
\epsilon_{s}=4.337\frac{1}{m_{a}R^{2}}-1.936\frac{Gm_{a}^{2}N}{R}-0.1214\frac{N|\lambda|}{m_{a}^{2}R^{3}} \, .
\end{equation}

Define dimensionless momentum $\kappa_{i}\equiv k_{i}R$, the energy conservation $\delta(\epsilon_{1}+\epsilon_{2}-\epsilon_{3}-\epsilon_{s})$ gives,
\begin{equation}\label{eq:kappa3}
\kappa_{3}^{2}=\kappa_{1}^{2}+\kappa_{2}^{2}+2m_{a}R^{2}|\epsilon_{s}| \, .
\end{equation}
The last term $2m_{a}R^{2}|\epsilon_{s}|$ is
\begin{equation}\label{eq:eps}
2m_{a}R^{2}|\epsilon_{s}|=\left|8.674-3.872\tilde{N}\tilde{R}-0.2428\frac{\tilde{N}}{\tilde{R}}\right| \, .
\end{equation}
This result will be used in the numerical calculation of $\alpha(\nu,\tilde{N})$. We note that for small $\tilde{N}$, the self-interaction can be neglected, and Eq.~(\ref{eq:radius}) gives $\tilde{N}\tilde{R}=2a/b$. Hence $2m_{a}R^{2}|\epsilon_{s}|=26.022$ for small $\tilde{N}$.

Next we calculate the dimensionless parameter $\beta$ in Eq.~(\ref{eq:1F1}). $\beta=GM_{s}m_{a}^{2}R$, and the gravitational potential energy of the axion star is 
\begin{equation}
U(r)=-\frac{GM_{s}m_{a}}{r}=-\frac{\beta}{m_{a}Rr} \, .
\end{equation}
We note from Eq.~(\ref{eq:dimless}) that 
\begin{equation}
\beta=\tilde{N}\tilde{R}=\frac{a+\sqrt{a^{2}-3bc\tilde{N}^{2}}}{b} \, .
\end{equation}
For small $\tilde{N}$, $\beta=2a/b=8.961$. We also define $\phi_{\vec{\kappa}}(\vec{\xi})=\sqrt{\kappa/(2\pi\beta)}\varphi_{\kappa}(\vec{\xi})$ and $\nu=k_{g}R$, and the mass growth rate $\Gamma_{\lambda}$ is
\begin{equation}\label{eq:gamlam}
\begin{split}
\Gamma_{\lambda}&=\frac{\lambda^{2}\rho_{g}^{2}}{16\pi^{2}m_{a}^{5}k_{g}^{2}}\nu^{-4}\frac{\beta^{3}}{c_{1}}\int\textrm{d}^{3}\vec{\kappa}_{1}\textrm{d}^{3}\vec{\kappa}_{2}\textrm{d}(\cos\theta_{3})d\varphi_{3}\, \\ & \left[e^{-\frac{1}{\nu^{2}}(\kappa_{1}^{2}+\kappa_{2}^{2})}-e^{-\frac{1}{\nu^{2}}(\kappa_{1}^{2}+\kappa_{3}^{2})}-e^{-\frac{1}{\nu^{2}}(\kappa_{2}^{2}+\kappa_{3}^{2})}\right]\cdot
\\
& (\kappa_{1}\kappa_{2})^{-1} \left|\int \textrm{d}^{3}\vec{\xi}\, (1+\xi^{2})^{-4} \phi_{\vec{\kappa}_{1}}(\vec{\xi}) \phi_{\vec{\kappa}_{2}}(\vec{\xi})\phi^{*}_{\vec{\kappa}_{3}}(\vec{\xi}) \right|^{2} \\
&\equiv \frac{\lambda^{2}\rho_{g}^{2}}{16\pi^{2}m_{a}^{5}k_{g}^{2}}\alpha(\nu,\tilde{N}) \, ,
\end{split}
\end{equation}
where $\alpha(\nu,\tilde{N})$ is a dimensionless numerical factor and $\kappa_{3}$ is given by Eq.~(\ref{eq:kappa3}) and Eq.~(\ref{eq:eps}).
For $\vec{\kappa}_{1}$ and $\vec{\kappa}_{2}$, we use the approximation 
\begin{equation}\label{eq:phika}
\phi_{\vec{\kappa}}(\vec{\xi})=J_{0}\left(2\sqrt{\beta\left(\xi-\vec{n}\cdot\vec{\xi}\right)}\right) \, .
\end{equation}
For $\vec{\kappa}_{3}$, the approximation can still be used with a slightly larger deviation because $\vec{\kappa}_{3}$ is generally larger, as can be seen from Eq.~(\ref{eq:kappa3}).

Note from Eq.~(\ref{eq:gamlam}) that the factor $\alpha(\nu,\tilde{N})$ only depends weakly on $\nu$ and $\tilde{N}$. The integration for $\kappa_{1}$ and $\kappa_{2}$ is dominated by $\kappa_{1},\kappa_{2}<\nu$. In this case the second line in Eq.~(\ref{eq:gamlam}) only has a slight dependence on $\kappa_{1}$ and $\kappa_{2}$. The approximation above shows that the integration in the third line is almost independent of $\kappa_{1}$ and $\kappa_{2}$. Hence the integration over $\kappa_{1}$ and $\kappa_{2}$ can be approximately performed, giving a result proportional to $k_{g}^{4}$. Because $\nu=k_{g}R$, we can see that $\alpha(\nu,\tilde{N})\sim \nu^{0}$. Furthermore, the $\tilde{N}$ dependence of $\alpha(\nu,\tilde{N})$ only enters through $\kappa_{3}$ which is much larger than $\kappa_{1}$ and $\kappa_{2}$. For small $\nu$ considered here, the second line in Eq.~(\ref{eq:gamlam}) is dominated by the first term. Hence we expect that $\alpha(\nu,\tilde{N})$ depends weakly on $\tilde{N}$. Numerical evaluation also corroborates these results.
We will take $\alpha(\nu,\tilde{N})\sim \mathcal{O}(60)$ as given by a numerical estimation.

We will also use the mass growth rate for gravitational interaction
\begin{equation}\label{eq:gammagravity}
\Gamma_{\rm gravity}=\frac{(4\pi G)^{2}m_{a}^{3}\rho_{g}^{2}}{k_{g}^{6}}\gamma(\nu) \, .
\end{equation}
In contrast to the case of self-interaction, the factor $\gamma(\nu)$ is proportional to $\nu^{4}$ for small $\nu$. An empirical fit from the simulation results Fig.10 from Ref.~\cite{Chan:2022bkz} gives an approximate expression for $\nu\lesssim 10.8$,
\begin{equation}\label{eq:gammanu}
\gamma(\nu)=\frac{0.21\nu^{4}}{1+0.036\nu^{4}} \, .
\end{equation}
Note that the definition of $\nu$ is slightly different here. For $\nu\gtrsim 10.8$, the axion star will be evaporated with $\gamma(\nu)<0$. However, for an axion star formed in an MC, $\nu<10.8$ and the axion star mass increases. We have verified that the value $\gamma(\nu)$ can also be obtained by the same method as we used for self-interaction, and the results approximately coincide with the simulation result Eq.~(\ref{eq:gammanu}).

The mass growth rate obtained in this section will be used to discuss the time required for an axion star to reach the maximum mass.

\section{Axion parameter dependence}
\label{sec:b}
We will discuss axion star mass growth and calculate the time that it reaches the maximum mass and becomes a bosenova. In the following, MCs refer to isolated miniclusters that are formed around matter-radiation equality. They form directly from spherical collapse of overdensed regions. MCHs refer to dark matter substructures formed in the late universe that result from disruption and merger of MCs. For relatively dense MCs, they may survive today as isolated dense MCs~\cite{Fairbairn:2017sil,Kavanagh:2020gcy}. We will discuss bosenova from boson stars condensed in isolated MCs and MCHs separately. 

The mass of the QCD axion depends on temperature for $T\gtrsim0.1\,{\rm GeV}$~\cite{Wantz:2009it},
\begin{equation}\label{eq:masstem}
m_{a}(T)=3.07\times10^{-9}\, {\rm eV}\left(\frac{f_{a}}{10^{12}\, {\rm GeV}}\right)^{-1}\left(\frac{T}{{\rm GeV}}\right)^{-3.34} \, .
\end{equation}
Using $m_{a}(T_{\rm osc})=3H(T_{\rm osc})$ we obtain the oscillating temperature for the QCD axion,
\begin{equation}
T_{\rm osc}=1.078\, {\rm GeV} \left(\frac{m_{a}}{10^{-5}\,{\rm eV}}\right)^{0.187} \, .
\end{equation}
Entropy conservation yields a constant value of $g_{*S}(T)R_{\rm scale}^{3}T^{3}$, where $g_{*S}$ is effective entropy degrees of freedom, we obtain the scale factor $R_{\rm scale}$ at oscillating temperature,
\begin{equation}
R_{\rm osc}=8.096\times 10^{-14} \left(\frac{m_{a}}{10^{-5}\,{\rm eV}}\right)^{-0.187} \, .
\end{equation}
where we take $R_{0}=1$ for the scale factor at present. Using  Eq.~(\ref{eq:m0}), we obtain the MC characteristic mass
\begin{equation}\label{eq:M0qcd}
\begin{split}
M_{0}&=\frac{4\pi^4}{3}\frac{\bar{\rho}_{a}(t_{0}) R_{\rm osc}^{-3}}{H^{3}(t_{\rm osc})} \\ & =1.235\times 10^{-9}\, M_{\odot} \left(\frac{m_{a}}{10^{-5}\,{\rm eV}}\right)^{-0.561} \, ,
\end{split}
\end{equation}
Here, we assumed that the dark matter is composed solely of axions. The MC mass may also depend on initial overdensity $\delta$. In the following, we take the MC mass used in Ref.~\cite{Visinelli:2018wza}, which gives
\begin{equation}\label{eq:MMC}
M_{\rm MC}= \frac{4\pi}{3}(1+\delta)\frac{\bar{\rho}_{a}(t_{0}) R_{\rm osc}^{-3}}{H^{3}(t_{\rm osc})}r_{0}=\frac{r_{0}}{\pi^{3}}(1+\delta)M_{0} \, .
\end{equation}
where a numerical factor $r_{0}$ of order unity is introduced.
The MC mass is proportional to $1+\delta$ because a region with an initial overdensity $\delta$ has a larger density $(1+\delta)\bar{\rho}_{a}(t_{\rm osc})$ at the oscillating temperature, resulting in a larger mass within the Hubble horizon.

The MC is assumed to have a NFW profile 
\begin{equation}\label{eq:NFW}
\rho(r)=\frac{\rho_{s}}{\frac{r}{r_{s}}(1+\frac{r}{r_{s}})^{2}} \, .
\end{equation}
The gas density $\rho_{g}$ appear in Eq.~(\ref{eq:gamlam}) and (\ref{eq:gammagravity}) is taken as the scale density $\rho_{s}$, whose value is taken as $\rho_{\rm mc}$ in Eq.~(\ref{eq:MCd}). This is a conservative estimate as the central density of MCs could be higher. The MC mass is taken as Eq.~(\ref{eq:MMC}) with $r_{0}=100$.
The virial radius $r_{v}$ is defined as the radius where the average density of the dark matter is $200\bar\rho_{a}(t_{0})$. The concentration is defined as $c=r_{v}/r_{s}$. The scale radius is
\begin{equation}\label{eq:rs}
r_{s}=\frac{1}{c}\left(\frac{3}{800\pi}\right)^{\frac{1}{3}}\left(\frac{M_{\rm MC}}{\bar\rho_{a}(t_{0})}\right)^{\frac{1}{3}}\, .
\end{equation}
The scale density can also be related to the MC mass $M_{\rm MC}$ and the concentration $c$,
\begin{equation}\label{eq:rhos}
\rho_{s}=\frac{M_{\rm MC}}{4\pi r_{s}^{3}\left[\ln(1+c)-\frac{c}{1+c}\right]}\, .
\end{equation}
The gas velocity $v_{g}$ is taken as the circular velocity $v_{c}$ at the scale radius $r_{s}$,
\begin{equation}\label{eq:vcc}
v_{c}(r_{s})=\sqrt{2\pi G\rho_{s}r_{s}^{2}(\ln 4-1)}\, .
\end{equation}
If we take $\rho_{s}=\rho_{\rm MC}(\delta)$ given by Eq.~(\ref{eq:MCd}),
and Eq.~(\ref{eq:rs}) and (\ref{eq:rhos}) yield a relation between $c$ and $\delta$,
\begin{equation}\label{eq:cdel}
\frac{200}{3}\frac{c^{3}}{\left[\ln(1+c)-\frac{c}{1+c}\right]}=140\delta^{3}(1+\delta)(1+z_{\rm eq})^{3}\, .
\end{equation}
For $\delta\gtrsim 1$, the solution $c(\delta)$ is well approximated by a power law,
$c(\delta)\approx 9.4\times 10^{3}\delta^{1.36}$.
These results will be used for calculations of boson star mass growth.

The maximum mass of the axion star can be calculated using $M_{s}=Nm_{a}=\tilde{N}/\sqrt{G|\lambda|}$, and take the maximum particle number $\tilde{N}=10.33$, 
\begin{equation}
M_{\rm max}=1.089\times 10^{-11}\,{M_{\odot}} \left(\frac{m_{a}}{10^{-5}\,{\rm eV}}\right)^{-2} \, .
\end{equation}
Here we take $\lambda=-0.346m_{a}^{2}/f_{a}^{2}$ and use the relation $m_{a}=5.7\times10^{-6}\,{\rm eV}(f_{a}/10^{12}\,{\rm GeV})^{-1}$ for the QCD axion.

We need to define an initial mass of axion star $M_{s0}$. The mass growth rate calculated in the last section can only be used after the axion star is formed in the MC. The axion stars within an MC must have the same properties (especially the same axion velocity) as isolated axion stars in the ground state. Initially, the axion star formed is not in an equilibrium state, and the axion velocity is close to the axion velocity $v_{g}$ in the MC, which is much larger than the axion velocity in an isolated axion star with the same mass. Because the mass growth rate depends sensitively on axion velocity, our calculation can only be used as an estimation before the axion star reaches a critical mass $M_{s0}$. For an isolated axion star, the axion velocity increases as it becomes heavier.  When the axion star reaches mass $M_{s0}$, the velocity of the axions in the MC equals the velocity of axions in an isolated axion star with mass $M_{s0}$. The critical mass $M_{s0}$ defined here is roughly equals to the characteristic mass at axion star formation defined in simulations~\cite{Levkov:2018kau}, $M_{s0}\approx c_{0}v_{g}/(Gm_{a})$ with $c_{0}\approx 3$.

\begin{equation}
\begin{split}
M_{s0}&=8.384\times 10^{-11}\,{M_{\odot}}\times \\ &\alpha_{0} \left[\delta^{3}(1+\delta)\right]^{\frac{1}{6}} \left(\frac{m_{a}}{10^{-5}\,{\rm eV}}\right)^{-1} \left(\frac{M_{0}}{M_{\odot}}\right)^{\frac{1}{3}} \, .
\end{split}
\end{equation}
For large $\delta$ and $m_{a}$, $M_{s0}$ may be larger than $M_{\rm max}$. We can only use our calculation to estimate the time required for the axion star to become a bosenova in such case.

\begin{figure}[t]
  \centering
  \includegraphics[width=8cm]{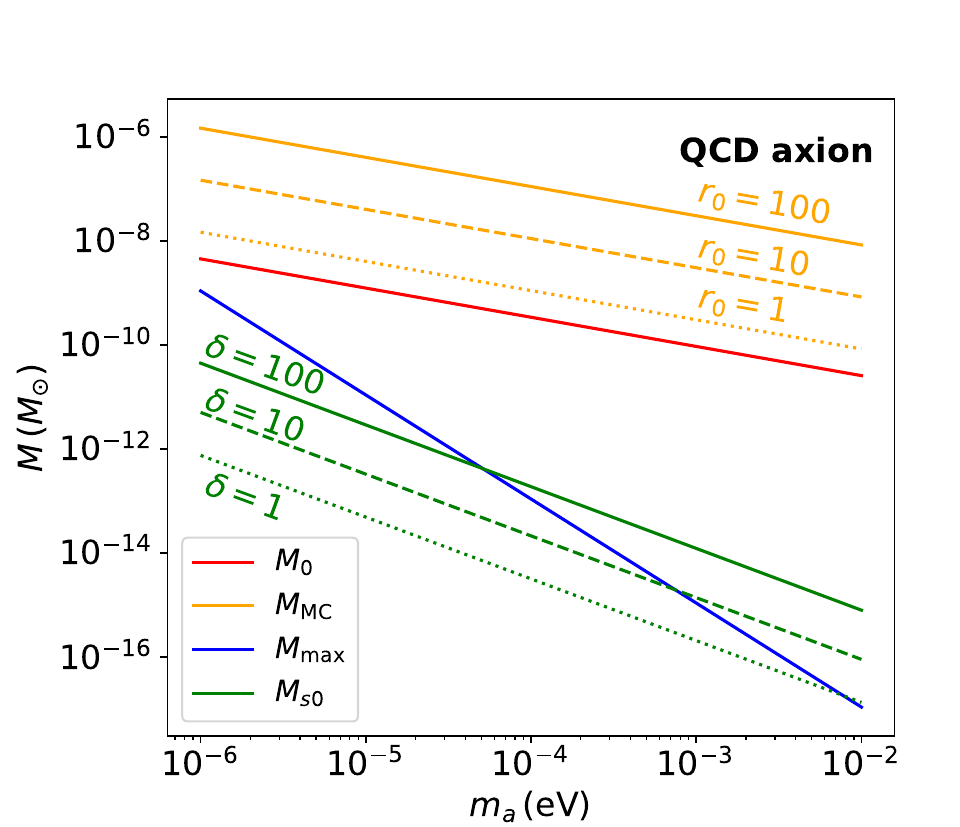}\\
  \caption{Comparison of the characteristic mass $M_{0}$ (\emph{red}), the MC mass Eq.~(\ref{eq:MMC}) with $\delta=100$ (\emph{orange}), the maximum mass of the dilute axion star $M_{\rm max}$ (\emph{blue}) and the initial mass of the axion star $M_{s0}$ (\emph{green}) for the QCD axion with different mass. We fix $\delta=100$ and take $r_{0}=100$, $r_{0}=10$ and $r_{0}=1$ for the solid, dashed and dotted orange line. We fix $r_{0}=100$ and take the overdensity $\delta=100$, $\delta=10$, $\delta=1$ for the solid, dashed and dotted green line.}
  \label{img1}
\end{figure}

In Fig.~\ref{img1}, we plot the MC characteristic mass $M_{0}$, the MC mass for $\delta=100$, the maximum mass of the boson star $M_{\rm max}$, and the axion star formation mass $M_{s0}$ for different QCD axion masses, shown as red, orange, green, and blue lines, respectively. As can be seen, for the QCD axion mass $m_{a}\sim 10^{-6}-10^{-2}\,{\rm eV}$, we have $M_{\rm max}<M_{0}$, which means that after a long enough time, the axion star in an MC will reach the maximum mass. For a larger axion mass, in case that $M_{s0}>M_{\rm max}$, the axion star explodes once it is formed. We estimate the time to become a bosenova as  $t_{c}\sim 2\tau_{c}=2/(\Gamma_{\rm gravity}+\Gamma_{\lambda})$ to use the simulation result in Ref.~\cite{Levkov:2018kau}, which gives an axion star mass $M_{s}=M_{s0}(t/\tau_{c}-1)^{1/2}$. During the first time duration $\tau_{c}$, the axion star does not form in the MC, but the axion velocity distribution evolves. The axion star spends an additional time $\tau_{c}$ to grow to mass $M_{s0}$. 

The estimation $t_{c}\sim 2\tau_{c}$ is a conservative estimate for bosenova explosion, because the axion star exceeds $M_{\rm max}$ at an earlier time. However, the axion velocity inside an MC is faster than that for an isolated axion star, 
the conclusion for the axion star upper mass limit $M_{\rm max}$ may not apply. For a smaller axion mass with $M_{s0}<M_{\rm max}$, we use the mass growth rate after $t_{c}$ and evolve the axion star mass until it reaches $M_{\rm max}$. 

The mass growth rate from gravitational interaction depends on $\nu$. For axion star mass equals to $M_{s0}$, using $M_{s0}=c_{0}v_{g}/Gm=Nm$, $\nu=mv_{g}R$ and Eq.~(\ref{eq:dimless}), we find $\nu=\tilde{N}\tilde{R}/c_{0}$. Eq.~(\ref{eq:radius}) gives the maximum value $\tilde{N}\tilde{R}\leq 2a/b$, hence we obtain $\nu\leq2.98$. For an axion star heavier than $M_{s0}$, the value $\nu$ is even smaller. Hence the requirement $\nu\lesssim 10.8$ is satisfied and we have $\Gamma_{\rm gravity}>0$. For $\nu\leq2.98$, we also have $\Gamma_{\lambda}>0$, hence the axion star will always grow after it reaches the mass $M_{s0}$. As the axion star grows, the radius $R$ decreases, and $\nu$ becomes smaller. Because $\alpha(\nu)\sim\nu^0$ and $\gamma(\nu)\sim\nu^4$ for small $\nu$,
the gravitational contribution to mass growth $\Gamma_{\rm gravity}$ decreases as the axion star grows, while $\Gamma_{\lambda}$ remains almost a constant.

For the QCD axion MC, the timescale of mass growth from gravitational interaction $t_{c,\,\rm gravity}$ for $\delta\sim 100$ is

\begin{equation}\label{eq:tgr}
\begin{split}
&t_{c,\,\rm gravity}=\frac{1}{\Gamma_{\rm gravity}}\\ &=1.41\,{\rm Gyr}\frac{\delta^{3}(1+\delta)^{3}/c^{6}(\delta)}{7.15\times10^{-29}} \frac{1}{\gamma(\nu)}\left(\frac{r_{0}}{\pi^{3}}\right)^{2} \left(\frac{m_{a}}{10^{-5}\,{\rm eV}}\right)^{1.878}\, .
\end{split}
\end{equation}
where the concentration $c(\delta)\approx 9.4\times 10^{3}\delta^{1.36}$.
The timescale of mass growth from self-interaction 
$t_{c,\lambda}$ for $\delta\sim 100$ is

\begin{equation}\label{eq:tlam}
\begin{split}
&t_{c,\, \lambda}=\frac{1}{\Gamma_{\lambda}}=3989 \,{\rm Gyr} \left[\frac{\delta^{3}(1+\delta)^{\frac{1}{3}}c^{2}(\delta)}{1.13\times10^{20}}\right]^{-1} \times \\ &\left(\frac{r_{0}}{\pi^{3}}\right)^{\frac{2}{3}} \left(\frac{\alpha(\nu,\tilde{N})}{60}\right)^{-1} \left(\frac{m_{a}}{10^{-5}\,{\rm eV}}\right)^{-1.374}\, ,
\end{split}
\end{equation}
as can be seen from Eq.~(\ref{eq:tgr}) and Eq.~(\ref{eq:tlam}), $t_{c,\,\rm gravity}$ and $t_{c,\, \lambda}$ are smaller than $13.7\,{\rm Gyr}$ for $\delta\gtrsim 100$. Self-interaction becomes important for larger axion masses.

\begin{figure}[t]
  \centering
  \includegraphics[width=8.5cm]{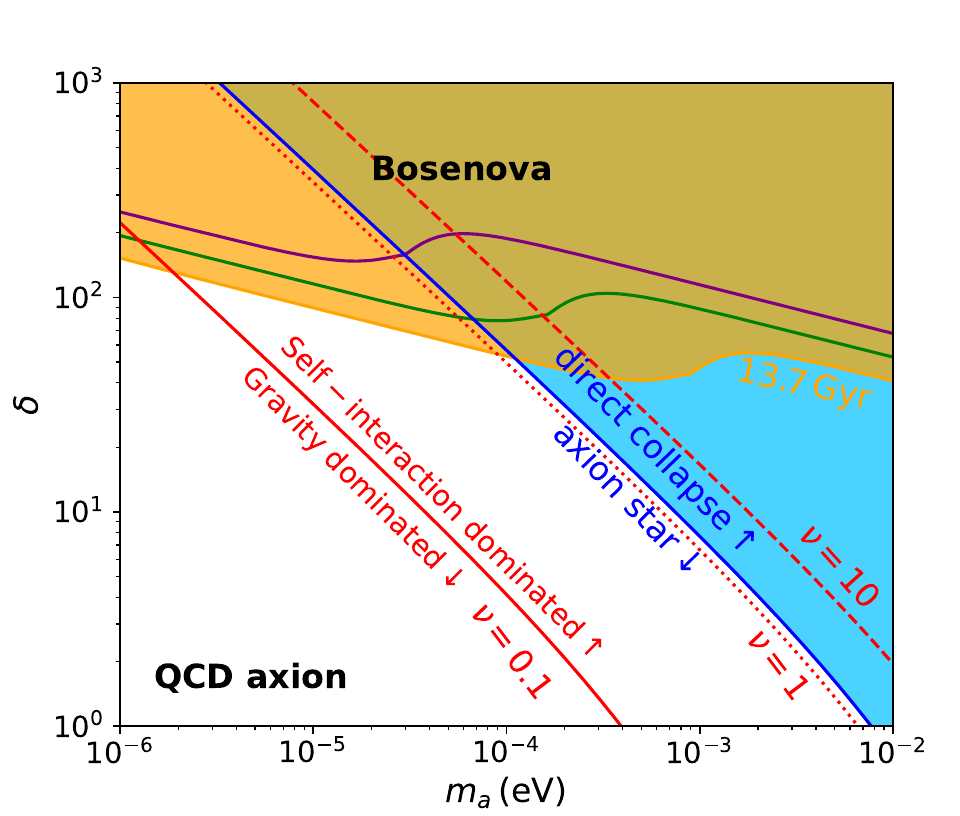}\\
  \caption{Parameter space for the axion MCs consisted of the QCD axion. The orange, green and purple lines represent the critical overdensity for bosenova. The MC mass is taken as Eq.~(\ref{eq:MMC}), and $r_{0}$ is taken as $1$, $10$, $100$, respectively. Bosenova occurs within $13.7\,{\rm Gyr}$ in the orange region for $r_{0}=1$. The MC density profile is taken as NFW profile. The right-hand side of the blue line shows the parameters that lead to $M_{s0}>M_{\rm max}$, hence the axion star becomes a bosenova once it is formed. The solid, dotted and dashed red lines are the boundary between gravity dominated and self-interaction dominated mass growth for $\nu=0.1$, $\nu=1$ and $\nu=10$. The left side of the red line is gravity dominated region, while self-interaction dominates its right side. }
  \label{img2}
\end{figure}
In Fig.~\ref{img2}, we plot the orange region where bosenova occurs within $13.7\,{\rm Gyr}$ for different initial overdensity $\delta$ and the QCD axion mass $m_{a}$. The orange, green and purple lines represent the critical overdensity for bosenova. Here we use the MC mass Eq.~(\ref{eq:MMC}), and $r_{0}$ is taken as $1$, $10$, $100$, respectively. For a larger initial overdensity $\delta$, the MC density is greater, hence it is faster for an axion star to reach the maximum mass. We note that bosenova occurs only for $\delta\gtrsim 100$. According to the simulations in Ref.~\cite{Vaquero:2018tib}, the probability for $\delta\gtrsim 100$ is roughly $10^{-7}$, which means that only $10^{-7}$ of the MCs will engender a bosenova.  The orange, green and purple lines on the large axion mass side are lowered due to axion self-interaction.
The solid, dotted, and dashed red lines show the boundary between gravity dominated and self-interaction dominated growth scenarios for $\nu=0.1$, $\nu=1$ and $\nu=10$.

\begin{figure}[t]
  \centering
  \includegraphics[width=8cm]{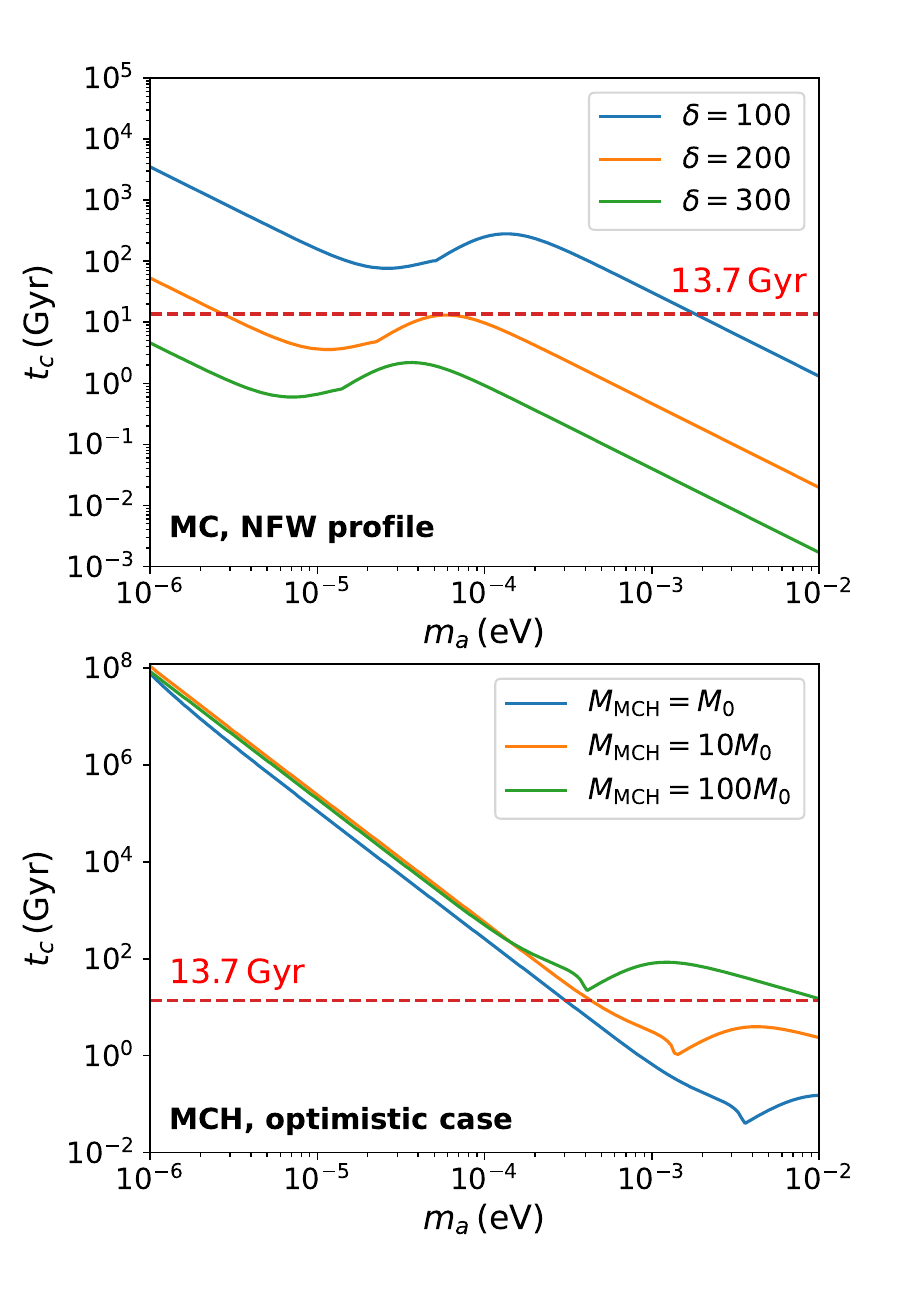}\\
  \caption{\emph{Top:} The time required to form a bosenova in a MC with NFW profile. The MC mass is taken as Eq.~(\ref{eq:MMC}) with $r_{0}=100$. The gas density $\rho_{g}$ in the mass growth rate is taken as the scale density $\rho_{s}$, given by $\rho_{\rm mc}$ in Eq.~(\ref{eq:MCd}). \emph{Bottom:} The time required to form a bosenova in a MCH in our optimistic scenario. The blue, orange and green lines shows the case for MCH mass $M_{\rm MCH}=M_{0}$, $M_{\rm MCH}=10M_{0}$ and $M_{\rm MCH}=100M_{0}$, respectively.}
  \label{imgc}
\end{figure}

In the top panel of Fig.~\ref{imgc}, we plot the time that boson star form and grow to become a bosenova. The blue, orange and green lines represent the time to become a bosenova for $\delta=100$, $\delta=200$ and $\delta=300$.

In the discussion above, we assume that MC is isolated once it is formed around the matter-radiation equality. For MCs with $\delta\gtrsim 100$, it is likely to survive and can be treated as isolated MCs~\cite{Fairbairn:2017sil,Kavanagh:2020gcy}. However, for most MCs with small $\delta$, they will inevitably experience mergers and tidal disruption, resulting in MCHs~\cite{Fairbairn:2017sil}. The boson star mass growth rate can also be used in the case of MCHs. Consider a MCH with a mass $M_{\rm MCH}$ and a NFW profile. The virial radius $r_{v}$ at matter-radiation equality is defined as the radius where the average density of the dark matter is $200\bar\rho_{a}(t_{\rm eq})$. The scale radius is
\begin{equation}
r_{s}=\frac{1}{c}\left(\frac{3}{800\pi}\right)^{\frac{1}{3}}\left(\frac{M_{\rm MCH}}{\bar\rho_{a}(t_{\rm eq})}\right)^{\frac{1}{3}}\, .
\end{equation}
where the concentration $c$, defined as the ratio of the virial radius $r_{v}$ to the scale radius $r_{s}$, is roughly $c\approx 4/\sqrt{M_{\rm MCH}/M_{0}}$~\cite{Xiao:2021nkb} at matter-radiation equality. The scale density of a MCH of mass $M_{\rm MCH}$ with concentration $c$ is
\begin{equation}
\rho_{s}=\frac{M_{\rm MCH}}{4\pi r_{s}^{3}\left[\ln(1+c)-\frac{c}{1+c}\right]}\, .
\end{equation}
After matter-radiation equality, the concentration increases while the scale radius and hence the characteristic MCH density does not change significantly~\cite{Xiao:2021nkb}. The MCH mass may increase but the majority of MCHs remains $M_{\rm MCH}\sim \mathcal{O}(1-100)M_{0}$.  
To estimate the mass growth rate, We take the gas velocity $v_{g}\approx v_{c}$ and the gas density $\rho_{g}\approx\rho_{s}$. For $M_{\rm MCH}=M_{0}$, Eq.~(\ref{eq:gamlam}) and (\ref{eq:gammagravity}) yield the boson star mass growth rate,
\begin{equation}
\Gamma_{\rm gravity}=2.745\times 10^{-22}\,{\rm s^{-1}}\gamma(\nu) \left(\frac{m_{a}}{10^{-5}\,{\rm eV}}\right)^{-1.878}\, .
\end{equation}
and the mass growth rate for self-interaction is
\begin{equation}
\Gamma_{\lambda}=8.182\times 10^{-33}\,{\rm s^{-1}}\alpha(\nu,\tilde{N}) \left(\frac{m_{a}}{10^{-5}\,{\rm eV}}\right)^{1.374}\, .
\end{equation}
The characteristic mass growth time is longer than the age of the Universe for $m_{a}\gtrsim 10^{-8}\,{\rm eV}$. 
This does not rule out the possibility of bosenova, because boson stars tend to form and grow in the center and the density near the center of a MCH is much larger than $\rho_{s}$. 

Next we discuss whether such boson stars may reach the maximum mass. We consider a MCH with a NFW profile. The density near the center of the MCH exhibits a spike $\rho(r)\propto r^{-1}$. The boson star will form near the center of MCH. In an optimistic scenario, the density spike is unaffected by tidal disruption and the boson star always stays in the density spike. As the boson star grows, the mass in the MCH will decrease. We assume that the MCH automatically evolves to maintain a NFW profile. To be more specific, when the MCH mass decreases from $M_{\rm MCH}$ to $M_{\rm MCH}'$, the density profile is taken as the NFW profile with scale density $\rho_{s}'=\rho_{s}M_{\rm MCH}'/M_{\rm MCH}$ and scale radius $r_{s}'=r_{s}$. Because the maximum mass of boson star $M_{\rm max}\ll M_{\rm MCH}$ for $m_{a}\gtrsim 10^{-6}\,{\rm eV}$, the decrease of MCH density is actually small as the boson star grows. The density $\rho_{g}$ in Eq.~(\ref{eq:gamlam}) and (\ref{eq:gammagravity}) is taken as the MCH density profile at the boson star radius. The characteristic velocity $v_{g}$ is taken as the axion circular velocity $v_{c}$ at the scale radius Eq.~(\ref{eq:vcc}). The change of $v_g$ during the boson star mass growth is neglected, because the total mass of MCH, including the boson star and the radius $r_{s}$ are treated unchanged. This should be reasonable at least for the case $M_{\rm max}\ll M_{\rm MCH}$.

Using our optimistic scenario, the time required for the boson star to reach the maximum mass is shown in the bottom panel of Fig.~\ref{imgc}. The blue, orange and green lines show the case for MCH mass $M_{\rm MCH}=M_{0}$, $M_{\rm MCH}=10M_{0}$ and $M_{\rm MCH}=100M_{0}$. It can be seen that bosenova may occur for the QCD axion mass $m_{a}\gtrsim 3\times10^{-4}\,{\rm eV}$. For a larger MCH mass, the MCH density is lower, hence the mass growth of the boson star becomes slower. The turnover points at the bottom of three lines correspond to  $M_{\rm max}=M_{s0}$. For a larger axion mass, the boson star collapses once it is formed.

We stress that the scenario requires that the boson star stays in the central spike of MCHs. The scenario also requires that the NFW profile of MCHs persists until the radius of the central boson star, with $r/r_{s}\lesssim 10^{-4}$. Otherwise, the axion density in the MCH is not enough for the boson star to grow.
\footnote{The axion star growth discussed in Ref.~\cite{Chang:2024fol} effectively takes the gas density $\rho_{g}=\rho_{s}$. In the optimistic scenario discussed here, the gas density is taken as the density of the spike in the center of MCHs, which can be as large as $10^{3}\rho_{s}$ to $10^{5}\rho_{s}$.} A detailed simulation that includes the density spike of the MCH is required for discussion of boson star grow in MCHs. The discussion here shows it is still possible that bosenova occurs in a MCH and the results should be regarded as an upper bound for bosenova from MCHs.

Isolated MCs with $\delta\sim 100$ are unlikely disrupted by stars and should survive today~\cite{Fairbairn:2017sil,Kavanagh:2020gcy,DSouza:2024flu}. The axion density in such axion MCs is larger than MCHs and tidal disruption is unlikely to occur. The main uncertainty when considering bosenova from isolated MCs is the fraction of axion MCs with $\delta\gtrsim 100$.

\section{Axion-like particles}
\label{sect:alp}

Next, we consider the case that MC is composed of ALPs. 
For an ALP from string theory, its mass can be taken as 
temperature-independent. The oscillating temperature is obtained from $m_{a}=3H(T_{\rm osc})$ as
\begin{equation}
T_{\rm osc}=\left(\frac{4}{5}\pi^{3}g_{*}\right)^{-\frac{1}{4}}(m_{a}M_{\rm pl})^{\frac{1}{2}}\, ,
\end{equation}
In this case, the MC characteristic mass is
\begin{equation}\label{eq:M02}
M_{0}=2.846\times 10^{-7}M_{\odot}\left(\frac{m_{a}}{10^{-10}\,{\rm eV}}\right)^{-\frac{3}{2}}\zeta(T_{\rm osc})\, ,
\end{equation}
where $T_{\rm osc}$ and $T_{0}$ are the temperature at $t_{\rm osc}$ and the cosmic microwave background temperature at present, respectively, while $\zeta\equiv g_{*S}(T_{\rm osc})g_{*}^{-\frac{3}{4}}(T_{\rm osc})/g_{*S}(T_{0})$ is a factor of order unity. $g_{*}$ and $g_{*S}$ are the number of effective relativistic degrees of freedom and effective entropy degrees of freedom, respectively, whose values are taken from Ref.~\cite{Laine:2015kra}. For the QCD axion, the oscillating temperature is insensitive to the axion mass, hence the value of $g_{*}$ and $g_{*S}$ are taken near $1\,{\rm GeV}$. For the ALP case, $T_{\rm osc}$ varies significantly with the axion mass, hence we keep the factor $\zeta(T_{\rm osc})$ in Eq.~(\ref{eq:M02}).

We still take the MC mass as $M_{\rm MC}=(r_{0}/\pi^{3})(1+\delta)M_{0}$.
The maximum mass is still given by $M_{s}=Nm_{a}=\tilde{N}/\sqrt{G|\lambda|}$ with $\tilde{N}=10.33$. We use the ALP self-interaction coupling $\lambda=-m_{a}^{2}/f_{a}^{2}$. The ALP mass $m_{a}$ and decay constant $f_{a}$ are treated as independent parameters. Hence the maximum mass is 
\begin{equation}
M_{\rm max}=1.124\times 10^{-6}\,{M_{\odot}} \left(\frac{m_{a}}{10^{-10}\,{\rm eV}}\right)^{-1} \left(\frac{f_{a}}{10^{12}\,{\rm GeV}}\right) \, .
\end{equation}
Note that Bosenova explosion requires $M_{\rm MC}>M_{\rm max}$, assuming no mergers between MCs. Hence an upper limit for $f_{a}$ is introduced by the requirement,
\begin{equation}
f_{a}<2.53\times 10^{11}\,{\rm GeV} \left(\frac{m_{a}}{10^{-10}\,{\rm eV}}\right)^{-\frac{1}{2}} \zeta \, .
\end{equation}
If we further assume that the ALP couples to photons and $f_{a}>10^{9}\,{\rm GeV}$, we find that bosenova explosion is only possible for ALP mass $m_{a}\lesssim 10^{-6}\,{\rm eV}$.

Here we still assume that the MC has a NFW profile, and the density $\rho_{g}$ and velocity $v_{g}$ are taken as $\rho_{s}$ and $v_{c}(r_{s})$, respectively. The concentration $c(\delta)$ can still be approximated by $c(\delta)\approx 9.4\times 10^{3}\delta^{1.36}$.
For axion stars composed of ALP, the characteristic time of mass growth from gravitational interaction for $\delta\sim \mathcal{O}(1)$ is

\begin{equation}\label{eq:tgr2}
\begin{split}
&t_{c,\,\rm gravity}=\frac{1}{\Gamma_{\rm gravity}}\\ &=1.206\times 10^{-5} \,{\rm Gyr} \left[\frac{\delta^{3}(1+\delta)^{3}/c^{6}(\delta)}{1.16\times 10^{-23}}\right] \left(\frac{r_{0}}{\pi^{3}}\right)^{2} \left(\frac{1}{\gamma(\nu)}\right) \zeta^{2}\, .
\end{split}
\end{equation}
We note that $t_{c,\,\rm gravity}$ is independent of ALP mass, and it is much shorter than the age of the Universe even for $\delta\sim \mathcal{O}(1)$. Hence we expect that bosenova occurs shortly after
the MC form at matter-radiation equality.

The characteristic time of mass growth from axion self-interaction is

\begin{equation}\label{eq:tlam_ALP}
\begin{split}
&t_{c,\, \lambda}=\frac{1}{\Gamma_{\lambda}}=172.6 \,{\rm Gyr}  \left[\frac{c^{2}(\delta)\delta^{3}(1+\delta)^{\frac{1}{3}}}{1.11\times 10^{8}}\right]^{-1} \times \\ &\left(\frac{\alpha(\nu,\tilde{N})}{60}\right)^{-1}  \left(\frac{m_{a}}{10^{-10}\,{\rm eV}}\right)^{2} \left(\frac{f_{a}}{10^{12}\,{\rm GeV}}\right)^{4} \left(\frac{r_{0}}{\pi^{3}}\right)^{\frac{2}{3}} \zeta^{\frac{2}{3}}\, .
\end{split}
\end{equation}
The effects of self-interaction become important for small $m_{a}$ and $f_{a}$ and larger $\delta$. The timescale of axion star mass growth is set by $t_{c}\sim\min(t_{c,\,\rm gravity},t_{c,\, \lambda})$, and $t_{c}\lesssim\mathcal{O}(10^{5})\, {\rm yr}$.

\begin{figure}[t]
  \centering
  \includegraphics[width=8.5cm]{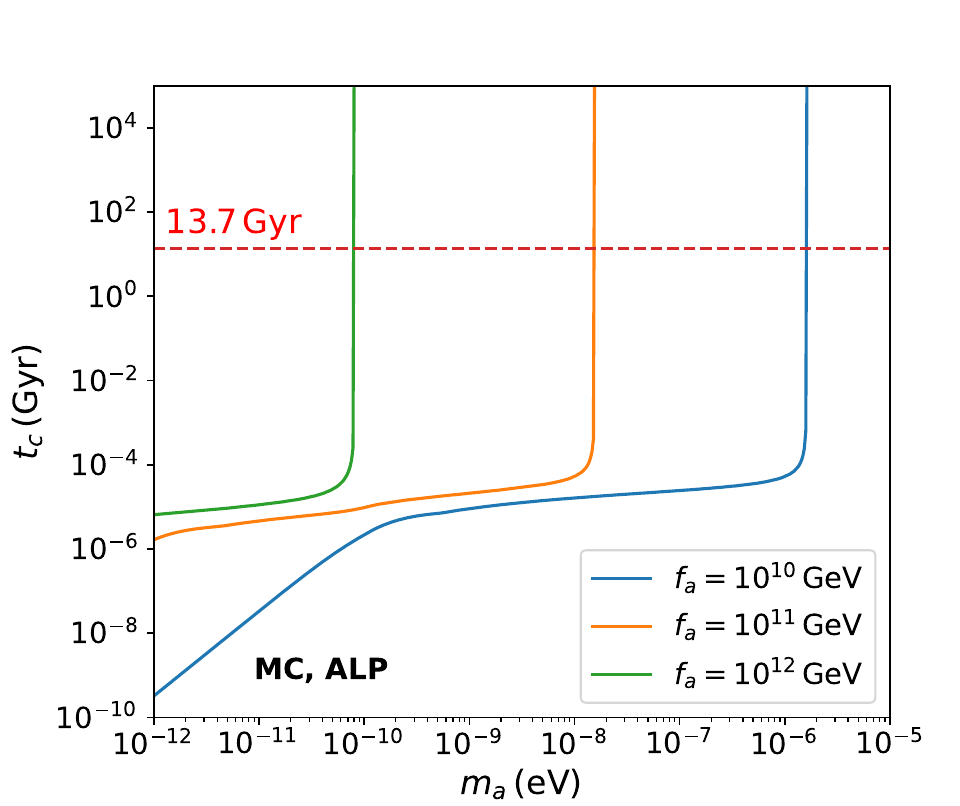}\\
  \caption{The time required to form a bosenova in an isolated ALP MC. Here we fix the initial overdensity $\delta=1$. The blue, orange and green lines are obtained for $f_a=10^{10}\,{\rm GeV}$, $f_a=10^{11}\,{\rm GeV}$ and $f_a=10^{12}\,{\rm GeV}$. The bosenova will never form if the MC mass is smaller than the maximum mass of the boson star, hence the solid lines become vertical on the right-hand side. }
  \label{imgy}
\end{figure}

Note that the overdensity $\delta$ in our discussions can not be arbitrarily large. This is because the MC density Eq.~(\ref{eq:MCd}) is obtained from pure gravitational collapse without considering the quantum pressure. For a large $\delta$, the MC density can be larger than the density of axion star with the same mass, which is unphysical because Eq.~(\ref{eq:MCd}) does not include the gradient pressure originated from the kinetic term of axion Lagrangian. The upper limit for $\delta$ can be estimate by equating the central density of the axion star $\rho_{c}=m_{a}|\chi(0)|^{2}=Nm_{a}/(c_{1}R^{3})$ to $\rho_{\rm mc}$, where the axion star mass $M=Nm_{a}$ is taken as the same value of MC mass. Here, if we take the self-interaction $\lambda=0$ so there is no maximum mass of the axion star, we obtain the maximum value of overdensity $\delta_{m}$ for MC composed of ALP and the QCD axion separately. For the ALP case, $\delta_{m}\approx 13$, while for the QCD axion, $\delta_{m}=3.07\times 10^{6}(m_{a}/10^{-5}\,{\rm eV})^{0.939}$. MCs with $\delta\gtrsim \delta_{m}$ still exist, but the density is not given by Eq.~(\ref{eq:MCd}).

\begin{figure}[t]
  \centering
  \includegraphics[width=8.5cm]{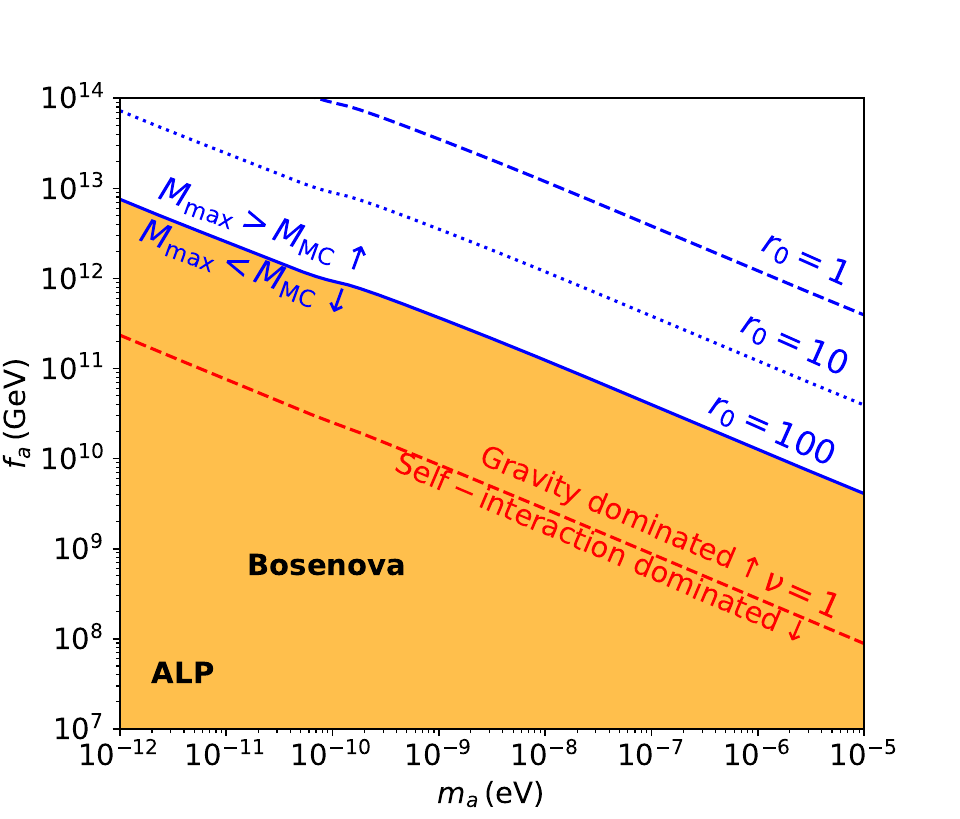}\\
  \caption{Parameter space for the ALP. The blue lines divide the region that $M_{\rm max}<M_{\rm MC}$ and $M_{\rm max}>M_{\rm MC}$. The blue solid, dotted and dashed lines use $r_{0}=100$, $r_{0}=10$ and $r_{0}=1$, corresponding to different MC mass. For the orange region with $M_{\rm max}<M_{\rm MC}$, bosenova will occur. The red lines divide the region that mass growth is dominated by gravity or self-interaction for $\nu=1$. Here we fix the MC initial overdensity $\delta=1$.}
  \label{img3}
\end{figure}

In Fig.~\ref{imgy}, the time for bosenova formation is plotted for different $f_{a}$. Here we take the overdensity $\delta=1$. If $M_{\rm MC}<M_{\rm max}$, bosenova will not form. For $M_{\rm MC}>M_{\rm max}$, bosenova formation time is much shorter than the age of the Universe, and should occur around matter-radiation equality when MCs are formed. In the ALP case, bosenova occurs before MCs aggregate to become MCHs, and we only need to consider bosenova from isolated MCs.

In Fig.~\ref{img3} we plot the ALP parameter space and shows the orange region with $M_{\rm max}<M_{\rm MC}$. In this region, the axion star will grow and exceed the maximum mass. The timescale is $t_{c}\sim\min(t_{c,\,\rm gravity},t_{c,\, \lambda})$, which is much shorter than the age of the Universe. Hence, we expect that ALP parameters within the orange region leads to bosenova explosion. Here we fix the MC initial overdensity $\delta=1$. For higher $\delta$, the mass growth process is even faster. The dashed red line represents the boundary line for gravity dominated and self-interaction dominated mass growth for $\nu=1$.


For the QCD axion, bosenova occurs for relatively dense isolated MCs with $\delta\gtrsim 100$. Because the uncertainties when discussing bosenova from MCHs, we only consider bosenova from isolated MCs resulting from spherical collapse when estimating bosenova event rate.  Only $10^{-7}$ of the isolated MCs have the density required for bosenova. We can estimate the event rate of bosenova within the Milky Way. The dark matter halo mass of the Milky Way is roughly $10^{12}\,M_{\odot}$~\cite{McMillan:2011wd}. For MC mass $M_{\rm MC}\sim 10^{-9}\, M_{\odot}$, and assuming that most of the axions are in the MCs, the number of MCs in the Milky Way is roughly $10^{21}$ and $10^{14}$ of them have $\delta\gtrsim 100$. Divided by the age of the Universe, the event rate of bosenova is roughly $10^{4}$ per year in the dark matter halo of the Milky Way. The uncertainties in the estimation include:

1. The fraction of MCs with $\delta\gtrsim 100$. This is the main uncertainty.

2. The fraction of the dark matter in the form of axion MCs or MCHs. Here we assume that all the axions are in the form of MCs or MCHs. 

3. The MC and MCH mass and the axion velocity near the center of the MC or MCH.

4. The mass and density of MCs with $\delta\gtrsim 100$ do not change significantly. Tidal disruption, accretion and merger of such MCs are unimportant. 

5. The final 'g' state of the absorption process $g+g\rightarrow g+s$ may escape from the MC due to a larger momentum. This effect should be small for the case that the maximum mass of the boson star is much smaller than the MC mass, which is true for $m_{a}\gtrsim 10^{-6}\,{\rm eV}$.  

6. The evolution of occupation number of axion gas states. We have assumed a Maxwellian momentum distribution Eq.~({\ref{eq:momentum}}) in our calculation. The distribution actually evolves as the boson star grow. The growth rate should only have a slight dependence on the form of distribution functions as long as the characteristic momentum $k_{g}$ unchanged~\cite{Levkov:2018kau,Dmitriev:2023ipv}. The evolution of the momentum distribution during the boson star growth need to be studied in numerical simulations, which may introduce some corrections.

For the ALP case, bosenova occurs for MCs with $\delta\sim 1$ at the time of photon decoupling. Because most MCs have $\delta\sim 1$, we expect that bosenova has a significant effect on the evolution of the Universe around photon decoupling. Especially, bosenova converts some of the nonrelativistic axions to mildly relativistic axions, which may contribute to the effective neutrino number $N_{\rm eff}$. Such effects deserve further studies.

\section{Summary}
\label{sect:diss}
In this work, we study the mass growth rate of an axion star after its formation within an axion MC or MCH. The axion star's mass increases as it continues to capture axions from the axion MC. Axion self-interaction plays an important role in the presence and the growth of the axion star. When the axion star reaches its maximum mass, instability leads to a bosenova. We treat the mass growth process as a quantum capture process, and derive the axion parameter dependence of the capture rate. Axion self-interaction may dominate the mass growth in a large fraction of the axion parameter space. 

For the QCD axion, bosenova occurs in the present Universe for MCs with $\delta\gtrsim 100$. For bosenova from MCs, the main uncertainty of the event rate is the fraction of MCs with $\delta\gtrsim 100$, which we take $10^{-7}$ here. For axion MCHs, bosenova is still possible for $m_{a}\gtrsim 3\times10^{-4}\,{\rm eV}$ in our optimistic scenario. However, bosenova from axion MCHs are still uncertain because the growth of boson star depends sensitively on the density of cusp structure near the center of MCHs, which may be affected by tidal effects. The results of MCHs should be regarded as an upper bound in the most optimistic case. Due to the large uncertainties, we do not include bosenovae from MCHs when estimating the event rate.  For the 
ALP, bosenova occurs roughly at the time of photon decoupling for MCs with $\delta\gtrsim 1$, which means that a significant fraction of MCs can generate bosenova.

For relatively large MC overdensities, the axion star growth time can be shorter than the cosmic age, as shown in Eq.~(\ref{eq:tlam}) and (\ref{eq:tlam_ALP}) for the QCD axion and ALP scenarios.  When bosenovae occur in our Universe, they emit relativistic axions and gravitational waves in the process, which may affect the effective radiation degrees of freedom $N_{\rm eff}$. If photonic coupling is large, electromagnetic radiation is also possible, potentially leading to new signals. Significant couplings to the Standard Model will also bring cosmic environmental effects into the axion star growth process, which will be of interest for further study.

\bigskip

\vspace{1cm}

{\bf Acknowledgements.} The authors acknowledge support by the National Natural Science Foundation of China under grant No. 12447105.

\bibliography{refs}
\end{document}